\documentclass[11pt]{article}
\usepackage{color}
\newcommand{\modif}[1]{\textcolor{black}{#1}}

\newcommand{\includegraphics}[1]{} 

\newcommand{\efface}[1]{}
\newcommand{\centre}[1]{\begin{center}#1\end{center}}
\newcommand{\NK}{Node-Kayles}
\newcommand{\SP}{Sprague-Grundy}
\newcommand{\nimsum}{\oplus}
\newcommand{\NN}{{\cal N}}
\newcommand{\PP}{{\cal P}}
\def\N{\mbox{I\hspace{-.15em}N}} 
\newcommand{\LL}{{\cal L}}

\newcommand{\floor}[1]{\left\lfloor~#1~\right\rfloor}
\newcommand{\ceil}[1]{\left\lceil~#1~\right\rceil}

\newtheorem{theorem}{Theorem}

\newtheorem{lemma}[theorem]{Lemma}
\newtheorem{corollary}[theorem]{Corollary}

\newcommand\qed{\mbox{}\hfill\rule{0.5em}{0.809em}\par\vskip 5mm}

\newenvironment{proof}[0]{\noindent\textbf{Proof.}}{\qed}

\newcommand{\modu}[1]{({\rm mod} #1)}

\newcommand{\keywords}[1]{\ \newline \noindent \textbf{Keywords: }#1}

\let\oldtitle=\title
\renewcommand{\title}[1]{\oldtitle{\Large{\textbf{#1}}}}

\newcommand{\institute}[1]{\newline \noindent
  \begin{minipage}{\textwidth}\begin{center}{\small #1}\end{center}\end{minipage}}

\newcommand{\email}[1]{{\small{\tt E-mail: #1}}}

\begin{document}

\title{Compound \NK\ on Paths}

\author{Adrien Guignard, \'Eric Sopena\\
  \email{{\small\tt \{Adrien.Guignard,Eric.Sopena\}@labri.fr}} \\ \ \\
  \institute{Universit\'e de Bordeaux \\
    LaBRI UMR 5800\\
    351, cours de la Lib\'eration \\
    F-33405 Talence Cedex, France}}

\date{\today}

\maketitle

\begin{abstract}
In his celebrated book {\em On \modif{Numbers} and Games} (Academic Press, New-York, 1976),
J.~H.~Conway introduced twelve versions of compound games. We analyze these twelve
versions for the \NK\ game on paths. For usual disjunctive compound, \NK\ has been
solved for a long time under normal play, while it is still unsolved under mis\`ere play.
We thus focus on the ten remaining versions, leaving only one of them unsolved.


\noindent\modif{{\bf AMS Mathematics Subject Classification 2000:} 91A46, 91A43.}
\end{abstract}

\section{Introduction}

An {\em impartial combinatorial game} involves two players, say $A$ and $B$,
who play alternately, $A$ having the first move, starting from some
starting position $G_0$~\cite{WW,ONAG}. When no confusion may arise, a game with starting
position $G_0$ is itself denoted by $G_0$. A {\em move} from a given
position $G$ consists in selecting the next position
within the finite set $O(G)=\{G_1,G_2,\dots,G_k\}$ of the {\em options} of $G$
($O(G)$ corresponds to the set of {\em legal moves} from $G$).
Such a game is {\em impartial} since the set $O(G)$ is the same for each player
playing on $G$ (otherwise, we speak about {\em partizan} games, that
we do not consider in this paper).
A common assumption is that the game finishes after a finite number of
moves and the result is a unique winner.
In {\em normal play}, the last player able to move (to a position
$G$ with $O(G)=\emptyset$) wins the game. Conversely, in {\em mis\`ere play},
the first player unable to move (from a position
$G$ with $O(G)=\emptyset$) wins the game.
A fundamental property of finite impartial combinatorial games is that
the {\em outcome} of any such game (that is which of the two
players has a winning strategy)
is completely determined by its starting position or, in other
words, by the game itself.

The main questions we consider when analyzing an impartial combinatorial game are $(i)$
to determine the outcome $o(G)$ of a game $G$ and $(ii)$ to determine which {\em strategy}
the winner has to use. 
We set $o(G)=\NN$ (resp. $o(G)=\PP$) when the first player (resp. second player), that is
the \underline{N}ext player (resp. the \underline{P}revious player), has a
winning strategy, and, in that case, $G$ is called a $\NN$-position
(resp. $\PP$-position).

For impartial combinatorial games under normal play, these questions 
can be answered using the \SP\ Theory~\cite{WW,ONAG}, independently discovered by 
Sprague~\cite{SPRAGUE-36} and Grundy~\cite{GRUNDY-39}: 
each game $G$ is equivalent to an instance of the game of Nim
on a heap of size $n$, for some $n\ge 0$. We then define the 
{\em \SP\ number} $\rho(G)$ of such a game $G$ by $\rho(G)=n$.
Therefore, in normal play, $o(G)=\PP$ if and only if $\rho(G)=0$.
For any game $G$, the value of $\rho(G)$ can be computed as the least
non negative integer which does not appear in the set
$\{\rho(G_i),\ G_i\in O(G)\}$, denoted by 
{\it mex\,}$(\{\rho(G_i),\ G_i\in O(G)\})$ (minimum excluded value).
The strategy is then the following: when playing on a game $G$ with
$o(G)=\NN$ (which implies $\rho(G)>0$), choose an option $G_i$ in
$O(G)$ with $\rho(G_i)=0$ (such an option exists by definition of $\rho$).

The \modif{{\em disjunctive sum}} of two impartial combinatorial games $G$ and
$H$, denoted by $G+H$, is the game inductively defined by
$O(G+H)=\{G_i+H,\ G_i\in O(G)\}$ $\cup$ $\{G+H_j,\ H_j\in O(H)\}$
(in other words, a move in $G+H$ consists in either playing on $G$
or playing on $H$).
The \SP\ value of $G+H$ is obtained as 
$\rho(G+H)=\rho(G)\nimsum\rho(H)$, where $\nimsum$ stands
for the binary XOR operation (called {\em Nim-sum} in this context).
The \modif{disjunctive} sum of combinatorial games
is the most common way of playing the so-called {\em compound games},
that is games made of several separated components. (The main subject
of this paper is to consider other ways of playing such compound games).

Following an inspiring paper by Smith~\cite{SMITH-66},
Conway proposed in~\cite[Chapter~14]{ONAG} twelve ways of playing compound games,
according to the rule deciding the end of the game, to the normal or
mis\`ere play, and to the possibility of playing on one or more components
during the same move.

\NK\ is an impartial combinatorial game played on
undirected graphs. A move consists 
in choosing a vertex and deleting this vertex
together with its neighbours. 
If we denote by $N^+(v)$ the set containing the vertex $v$ together with
its neigbours, we then have $O(G)=\{G\setminus N^+(v),\ v\in V(G)\}$
for every graph (or, equivalently, game) $G$. 
If $G$ is a non-connected graph with $k$ components, say $C_1$, $C_2$, $\dots$, $C_k$,
playing on $G$ is equivalent to playing on the \modif{disjunctive} sum
$C_1+C_2+\dots +C_k$ of its components (since a move consists in choosing
a vertex in exactly one of the components of $G$).

\NK\ is a generalisation of Kayles~\cite[Chapter~4]{WW}, independently introduced
by Dudeney~\cite{DUDENEY-10} and Loyd~\cite{LOYD-14}. 
This original game is played on a row of pins by two skilful
players who could knock down either one or two adjacent pins. 

Playing \NK\ on a path is equivalent
to a particular {\em Take-and-Break} game introduced by
Dawson~\cite{DAWSON-35}, and now known as {\em Dawson's chess}, which
corresponds to the octal game {\bf 0.137} (see~\cite[Chapter~4]{WW},
 \cite[Chapter~11]{ONAG}, or~\cite{WEB-FLAMMENKAMP} for more details).
This game has been completely solved by using \SP\ Theory (see
Section~\ref{sec:disjunctive}).

\NK\ has been considered by several authors.
Schaeffer~\cite{SCHAEFFER-78} proved that deciding the outcome of
\NK\ is PSPACE-complete for general graphs. In~\cite{BODLAENDER-KRATSCH-02},
Bodlaender and Kratsch proved that this question is polynomial time solvable
for graphs with bounded asteroidal number. (This class contains several
well-known graph classes such as cographs,
cocomparability graphs or interval graphs for instance.)
Bodlaender and Kratsch proposed the problem of determining the complexity
of \NK\ on trees. To our best knowledge, this problem is still unsolved.
In 1978 already, Schaeffer mentionned as an open problem to determine
the complexity of \NK\ on {\em stars}, that is trees having exactly one
vertex of degree at least three. Fleischer and Trippen proved
in~\cite{FLEISCHER-TRIPPEN-04}
that this problem is polynomial time solvable.

In this paper, we investigate Conway's twelve versions of compound games for
\NK\ on paths. Let $P_n$ denote the path with $n$ vertices and, for any $i$ and $j$,
$P_i\cup P_j$ denote the {\em disjoint union} of $P_i$ and $P_j$. As observed before, we
have $O(P_1)=O(P_2)=P_0$,
$O(P_3)=\{P_0,P_1\}$
and $O(P_n)=\{P_{n-2},P_{n-3}\}\cup\{P_i\cup P_j,\ j\ge i\ge 1,\ i+j=n-3\}$
(and, of course, $O(P_0)=\emptyset$).
With initial position $P_n$, any further position will thus be made of
$k$ disjoint paths, $P_{i_1}\cup P_{i_2}\cup\dots\cup P_{i_k}$,
with $i_1+i_2+\dots +i_k\le n-3(k-1)$ (since the only way to break
a path into two separated paths is to delete three ``non-extremal'' vertices),
which corresponds to a compound game.
Different rules for playing on this set of paths will lead to (very)
different situations.

This paper is organised as follows. In Section~\ref{sec:conway}, we present
in more details \modif{Conway's twelve versions} of compound games 
together with the tools
available for analyzing them,
\modif{as introduced in Conways's book~\cite[Chapter~14]{ONAG}}. 
We then consider these twelve versions 
of \NK\ on paths in Section~\ref{sec:variations} and discuss some 
possible extensions
in Section~\ref{sec:discussion}.

\section{Conway's twelve versions of compound games}
\label{sec:conway}

We recall in this section the twelve versions of compound games
introduced by Conway~\cite[Chapter~14]{ONAG}. 
Let $G$ be a game
made of several independent games $G_1$, $G_2$, $\dots$, $G_k$
(imagine for instance that we are playing \NK\ on a graph
$G$ with connected components $G_1$, $G_2$, $\dots$, $G_k$).
As we have seen in the previous section, 
the game $G=G_1+G_2+\dots +G_k$ is the {\em disjunctive compound} game obtained
as the disjunctive sum of its components. In this situation,
a {\em compound move} consists in making one legal move in
exactly one of the components.
By modifying this moving rule, we define a {\em conjunctive
compound} game (a move consists in playing in {\em all}
components \modif{simultaneously}) and a {\em selective compound}
game (a move consists in playing in any number $\ell$ of components,
$1\le\ell\le k$).

We can also distinguish two rules for ending such a compound game:
the game ends either when {\em all} the components have ended 
({\em long rule}) or as soon as one of the components has ended
({\em short rule}).

Finally, we have already seen that there are two different ways of deciding
who is the winner of a game, according to the {\em normal}
or {\em mis\`ere} rule.

Combining these different rules, we get twelve different versions of
compound games. Considering that the long rule is more natural for
selective and conjunctive compounds, while the short rule is more
natural for conjunctive compound, Conway proposed the following
terminology:

\begin{tabular}{rl}
  \\
  {\em disjunctive compound} & long ending rule, normal or mis\`ere play \\
  {\em diminished disjunctive compound} & short ending rule, normal or mis\`ere play \\
  {\em conjunctive compound} & short ending rule, normal or mis\`ere play \\
  {\em continued conjunctive compound} & long ending rule, normal or mis\`ere play \\
  {\em selective compound} & long ending rule, normal or mis\`ere play \\
  {\em shortened selective compound} & short ending rule, normal or mis\`ere play \\
  \\
\end{tabular}

We now recall how one can determine the outcome of these various compound games
\modif{(more details can be found in~\cite[Chapter~9]{WW} for conjunctive compounds and
in~\cite[Chapter~10]{WW} for selective compounds)}.

\vskip 4mm

\noindent{\bf Disjunctive compound.}
Under normal play, the main tool is the \SP\ Theory introduced in the previous section.
The {\em normal} \SP\ number $\rho(G)$ is computed as the Nim-sum
$\rho(G_1)\nimsum\rho(G_2)\nimsum\dots\nimsum\rho(G_k)$
(with $\rho(E)=0$ for any ended position $E$)
and $o(G)=\PP$ if and only if $\rho(G)=0$.

The situation for mis\`ere play is more complicated and the most useful
features of the \SP\ Theory for normal play have no natural
counterpart in mis\`ere play~\cite[Chapter 13]{WW}.
For instance, Kayles has been solved under normal play
in 1956, independently by Guy and Smith~\cite{GUY-SMITH-56}
and by Adams and Benson~\cite{ADAMS-BENSON-56} (the
\SP\ sequence has a period of length 12 after a preperiod
of length 70) while a solution of Kayles under mis\`ere play
was only given by Sibert in 1973 (and published in
1992~\cite{SIBERT-CONWAY-92}).
Three main approches have been used in the \modif{literature} to solve
mis\`ere impartial games:
{\em genus theory}~\cite{ALLEMANG-01,WW},
{\em Sibert-Conway decomposition}~\cite{SIBERT-CONWAY-92}
and {\em mis\`ere quotient semigroup}~\cite{PLAMBECK-05}.
These techniques cannot be summarized in a few lines and,
since we will not use them in this paper, we refer the
interested reader to the corresponding references
(see also~\cite{WEB-PLAMBECK}).

\vskip 4mm

\noindent{\bf Diminished disjunctive compound.}
Under both normal and mis\`ere play, we use the {\em foreclosed \SP\ number},
denoted by $F^+(G)$ (resp. $F^-(G)$) in normal (resp. mis\`ere) play, and
defined as follows.
Let us declare a position to be {\em illegal} if the game has just ended
or can be ended in a single {\em winning} move (note here that winning moves are
not the same under normal and mis\`ere play). 
If a position is illegal, its foreclosed \SP\ number
is {\em undefined}, otherwise its foreclosed \SP\ number
is simply its usual \SP\ number.
The foreclosed \SP\ number
of $G$ is then defined if and only if those of 
$G_1$, $G_2$, $\dots$, $G_k$ are all defined and, in that case, is computed
as their Nim-sum.
Now, the outcome of $G$ is $\PP$ if its foreclosed \SP\ number is 0
or some component has outcome $\PP$ but undefined foreclosed \SP\ number.

\vskip 4mm

\noindent{\bf Conjunctive compound.}
\modif{
In that case, the game ends as soon as one of the components ends.
Therefore, ``small'' components (that can be ended in a small
number of moves) must be played carefully:
a player has interest in winning quickly on winning components
and postponing defeat as long as possible on losing ones.
Considering this strategy, a game lasts
for a number of moves than can be easily computed. This number
of moves is called the {\em remoteness} of the game.
}
Under normal play, the remoteness $R^+(G)$ is computed as follows:
(i) if $G$ has an option of even remoteness, $R^+(G)$ is one more the
{\em minimal even} remoteness of any option of $G$,
(ii) if not, the remoteness of $G$ is one more than the {\em maximal odd}
remoteness of any option of $G$. Moreover, the remoteness of an
ended position is 0. A game $G$ will then have outcome $\PP$ if and
only if $R^+(G)$ is {\em even}
\modif{(the second player will play the last move)}.

Under mis\`ere play, the remoteness $R^-(G)$ is computed
similarly, except that we interchange the words {\em odd} and {\em even}
in the above rules. A game $G$ will now have outcome $\PP$ if and
only if $R^-(G)$ is {\em odd}.

\vskip 4mm

\noindent{\bf Continued conjunctive compound.}
\modif{
Now, the best strategy is to win slowly on winning components
and to lose quickly on losing components. The number of moves
of a game under such a strategy is called}
the {\em suspense} number of a game, denoted either
$S^+(G)$ or $S^-(G)$. The rules for computing this number in normal
play are the following:
(i) if $G$ has an option of even suspense number, $S^+(G)$ is one more the
{\em maximal even} suspense number of any option of $G$,
(ii) if not, the suspense number of $G$ is one more than the {\em minimal odd}
suspense number of any option of $G$. Moreover, the suspense number of an
ended position is 0. 
As before, for computing the suspense number under mis\`ere play,
we interchange the words {\em odd} and {\em even}
in the above rules.

A game $G$ will have outcome $\PP$ under normal play (resp. mis\`ere play)
if and
only if $S^+(G)$ is {\em odd} (resp. $S^-(G)$ is {\em even}).

\vskip 4mm

\noindent{\bf Selective compound.}
\modif{
The strategy here is quite obvious: to win the game
under normal play, a player
has to play on all winning components. Therefore,}
the outcome of $G$ is $\PP$ if and only if the
outcomes of $G_1$, $G_2$, $\dots$, $G_k$ are all $\PP$.
Under mis\`ere play, 
\modif{the winning strategy is the same, except when
all the remaining components are losing. If there is only one
such component, the player will lose the game. Otherwise, he can
win the game by playing on all but one of these losing components.
Therefore,}
unless all but one of the components of $G$ have ended,
the outcome of $G$ is the same as in normal play. Otherwise,
its outcome is $\PP$ if and only if the outcome of the only
remaining component is $\PP$.

\vskip 4mm

\noindent{\bf Shortened selective compound.}
\modif{Again, to win the game, a player
has to play on all winning components. But when
all components are losing, the player will lose
the game (even under mis\`ere play, since he will
necessary reach some configuration in which he cannot
play on all but one component without ending one
of these components).
Hence,
the} rule here is even simpler than the previous one:
under both normal play and mis\`ere play, 
the outcome of $G$ is $\PP$ if and only if the
outcomes of $G_1$, $G_2$, $\dots$, $G_k$ are all $\PP$.
Note that under normal play, all positions have the same outcome
in selective compound and in shortened selective compound.

\section{Compound \NK\ on paths}
\label{sec:variations}

Recall that for every path $P_n$ of order $n\ge 3$, the set of
options of $P_n$ in \NK\ is given by
\begin{equation}
O(P_n)=\{P_{n-2},P_{n-3}\}\cup\{P_i\cup P_j,\ 
j\ge i\ge 1,\ i+j=n-3\}.
\label{eq:options_Pn}
\end{equation}
In this section, we recall what is known for the usual disjunctive
compound \NK\
and analyze the ten other versions of compound \NK\ introduced in
the previous section. In each case, we will first try to characterize
the set $\LL=\{i\in\N,\ o(P_i)=\PP\}$ of {\em losing paths}
and then consider the complexity of determining the
outcome of any position (disjoint union of paths).
Finally, we will study the complexity of the {\em winning strategy} which
consists in finding, for any position 
with outcome
$\NN$, an option with outcome $\PP$.

\subsection{Disjunctive compound}
\label{sec:disjunctive}

Disjunctive composition is the most common way of considering
compound games. We recall here what is known (and unknown) for
disjunctive compound \NK\ on paths.

\vskip 4mm

\begin{center}{\sc Normal play}\end{center}
This game has been solved using the \SP\ Theory~\cite[Chapter 4]{WW}.
The sequence $\rho(P_0)\rho(P_1)\rho(P_2)\ldots\rho(P_{n-1})\rho(P_n)\ldots\ $
is called the {\em \SP\ sequence} of \NK.
It turns out that this sequence is {\em periodic},
with period 34, after a preperiod of size 51.
We then have:
$$\begin{array}{rcl}
\LL & = & \{0,4,8,14,19,24,28,34,38,42\}\\
     & & \cup\ \{54+34i,58+34i,62+34i,72+34i,76+34i,\ i\ge 0\}
     \end{array}
     $$
Determining the outcome of a path can thus be done in constant time.
For a disjoint union of paths, we need to compute the Nim-sum of the
\SP\ numbers of its components, which can be done in linear time.
Let now $G=P_{i_1}\cup P_{i_2}\cup\dots\cup P_{i_\ell}$ be any $\NN$-position
and assume $\rho(P_{i_1})\le\rho(P_{i_2})\le\dots\le\rho(P_{i_\ell})$.
Let $i_j\in\{1,2,\dots,\ell\}$ be the largest index such that
$(i)$ the number of components with \SP\ number $\rho(P_{i_j})$ is
odd and $(ii)$ for every $r>\rho(P_{i_j})$,  
the number of components with \SP\ number $r$ is even.
Thanks to the properties of the operator $\nimsum$, we have
$\rho(P_{i_j})>\nimsum_{k\in\{1,\dots,\ell\}\setminus\{j\}}\{P_{i_k}\}$.
Therefore, by choosing an option $H$ of $P_{i_j}$ with
$\rho(H)=\nimsum_{k\in\{1,\dots,\ell\}\setminus\{j\}}\{P_{i_k}\}$,
we get an option of $G$ with \SP\ number 0.
Such a ``winning move'' can thus be found in linear time. 

\vskip 4mm

\begin{center}{\sc Mis\`ere play}\end{center}

On the other hand, the problem is still open for \NK\ on
paths under mis\`ere play~\cite[Chapter 13]{WW}.

\subsection{Diminished disjunctive compound}
\label{ss:dimdis}

Recall that in this version of disjunctive compound, the game ends as soon
as one of the components has ended.

We shall compute the foreclosed \SP\ number of paths.
Under normal play, we shall prove that the corresponding sequence
is periodic and that the set of losing positions is finite.
On the other hand, we are unable to characterize the set of losing
positions under mis\`ere play.

\vskip 4mm

\centre{{\sc Normal play}}

Recall that the foreclosed \SP\ number of illegal positions (that is ended positions
or positions that can be won in one move) is undefined. Hence, we will note
$F^+(P_0)=F^+(P_1)=F^+(P_2)=F^+(P_3)=*$.
The foreclosed \SP\ number of other positions is computed as the usual \SP\ number,
using the $mex$ operator.
Hence, from~(\ref{eq:options_Pn}), we get for every $n\ge 4$:
$$F^+(P_n)=mex(\{F^+(P_{n-2}),F^+(P_{n-3})\}\cup
\{F^+(P_i\cup P_j),\ j\ge i\ge 1,\ i+j=n-3\}),$$
with $F^+(P_i\cup P_j)=F^+(P_i)\nimsum F^+(P_j)$.

Using that formula, 
\modif{and the fact that
$x\nimsum *=*\nimsum x=*$ for every $x$},
 we can compute the {\em foreclosed \SP\ sequence},
given as 
$F^+(P_0) F^+(P_1) F^+(P_2) \dots F^+(P_{n-1}) F^+(P_n) \dots\ $

In~\cite{GUY-SMITH-56}, Guy and Smith proved a useful {\em periodicity
theorem} for octal games (recall that \NK\ on paths is the octal game
{\bf 0.137}), which allows to ensure the periodicity of the usual
\SP\ sequence whenever two occurrences of the period have been computed.
This theorem can easily be extended to the foreclosed \SP\ sequence
in our context and we have:

\begin{theorem}
Suppose that for some $p>0$ and $q>0$ we have
$$F^+(P_{n+p})=F^+(P_n)\ \mbox{for every $n$ with $q\le n\le 2q+p+2$.}$$
Then
$$F^+(P_{n+p})=F^+(P_n)\ \mbox{for every $n\ge q$.}$$
\label{th:ddc_normal}
\end{theorem}

\begin{proof}
We proceed by induction on $n$. If $n\leq 2q+p+2$, the equality holds.
Assume now that $n\geq 2q+p+3$. Recall that
$$O(P_{n+p})=\{P_{n+p-2},P_{n+p-3}\} \cup
    \{P_i\cup P_j,\ j\ge i\ge 1,\ i+j=n+p-3\}.$$
Hence, we have
$$\begin{array}{rl}
F^+(P_{n+p})=mex\ ( & \{F^+(P_{n+p-2}),F^+(P_{n+p-3})\}\\
 & \cup\ \{F^+(P_i)\nimsum F^+(P_j),\ j\ge i\ge 1,\ i+j=n+p-3\}\ ).
 \end{array}$$

Since $n-2<n$ and $n-3<n$, we get by induction hypothesis
$F^+(P_{n-2})=F^+(P_{n+p-2})$ and
$F^+(P_{n-3})=F^+(P_{n+p-3})$.
Similarly, since $q+p\le\floor{\frac{n+p-3}{2}}-p\le j-p<n-3$, we get 
$F^+(P_{j-p})=F^+(P_j)$ and thus 
$F^+(P_{n+p})=F^+(P_n)$.
\end{proof}

By computing the foreclosed \SP\ sequence, we find a finite number of losing positions and,
thanks to Theorem~\ref{th:ddc_normal}, we get that this sequence is periodic, with period 84, after
a preperiod of length 245 (see Table~\ref{table:foreclosed}, the period is underlined).

\begin{table}
$$\begin{array}{|c|ccccl|}
\hline
n & \multicolumn{5}{c|}{F^+(P_n)} \\
\hline
0-49 & \mbox{\small ****}001120 & 0112031122 & 3112334105 & 3415534255 & 3225532255 \\
50-99 & 0225042253 & 4423344253 & 4455341553 & 4285322853 & 4285442804 \\
100-149 & 4283442234 & 4253345533 & 1253322533 & 2253422534 & 2253422334 \\
150-199 & 2233425334 & 4533425532 & 2553425544 & 2554425344 & 2234425334 \\
200-249 & 5533125342 & 2533225342 & 2534225342 & 2334223342 & 53344\underline{53342}  \\
250-299 & \underline{5532255342} & \underline{5344255442} & \underline{5344253442} & 
   \underline{5334553342} & \underline{5342253322} \\
300-349 & \underline{5342253422} & \underline{5342233422} & 
    \underline{334253342}5 & 3342553225 & \dots \\
\hline
\end{array}$$
\caption{\label{table:foreclosed}
The foreclosed \SP\ sequence under normal play}
\end{table}

Hence we have:

\begin{corollary}
$\LL=\{0,4,5,9,10,14,28,50,54,98\}$.
\end{corollary}

Determining the outcome of any disjoint union of paths or finding
a winning move from any $\NN$-position can be done in linear time,
using the same technique as in the previous subsection.

\vskip 4mm

\centre{{\sc Mis\`ere play}}

In that case, we have $F^-(P_0)=*$,
$F^-(P_1)=F^-(P_2)=0$,
$F^-(P_3)=F^-(P_4)=1$ and, for every $n\ge 5$:
$$F^-(P_n)=mex(\{F^-(P_{n-2}),F^-(P_{n-3})\}\cup
\{F^-(P_i\cup P_j),\ j\ge i\ge 1,\ i+j=n-3\}),$$
with $F^-(P_i\cup P_j)=F^-(P_i)\nimsum F^-(P_j)$.

Using that formula, \modif{and the fact that
$x\nimsum *=*\nimsum x=x$ for every $x$},
we have computed the mis\`ere foreclosed
\SP\ number of paths up to $n=10^6$, without being able to
discover any period. Some statistics on the corresponding
sequence are summarized in Table~\ref{table:misere}, where:
\efface{
\begin{itemize}
\item $n$ is the upper bound of the considered interval $I=[1,n]$,
\item $NbZ$ is the number of paths in $I$ with foreclosed \SP\ number 0,
\item $Max$ is the maximal foreclosed \SP\ number on $I$,
\item $Mean$ is the mean of the foreclosed \SP\ numbers on $I$,
\item $Deviation$ is the standard deviation of the foreclosed \SP\ numbers on $I$,
\item $FreqV$ is the most frequently encountered foreclosed \SP\ number on $I$,
\item $\% FreqV$ is the percentage of apparition of FreqV on $I$,
\item $MaxZ$ is the largest index of a path in $I$ with foreclosed \SP\ number 0,
\item $PosMax$ is the index of the largest foreclosed \SP\ number on $I$.\\
\end{itemize}
}

$-$ $n$ is the upper bound of the considered interval $I=[1,n]$,

$-$ $NbZ$ is the number of paths in $I$ with foreclosed \SP\ number 0,

$-$ $Max$ is the maximal foreclosed \SP\ number on $I$,

$-$ $Mean$ is the mean of the foreclosed \SP\ numbers on $I$,

$-$ $Deviation$ is the standard deviation of the foreclosed \SP\ numbers on $I$,

$-$ $FreqV$ is the most frequently encountered foreclosed \SP\ number on $I$,

$-$ $\% FreqV$ is the percentage of apparition of FreqV on $I$,

$-$ $MaxZ$ is the largest index of a path in $I$ with foreclosed \SP\ number 0,

$-$ $PosMax$ is the index of the largest foreclosed \SP\ number on $I$.

\begin{table}
$$
\begin{tabular}{|c|c|c|c|c|c|c|c|c|}
\hline
$n$ & $NbZ$ & $Max$ & $Mean$ & $Deviation$ & $FreqV$ & $\% FreqV$ & $MaxZ$ & $PosMax$\\
\hline
10 & 3 & 4 & 1.4 & 1.08 & 0 & 30\% & 8 & 9\\ \hline
$10^2$ & 8 & 11 & 4.23 & 2.4114 & 2 & 15\% & 98 & 61\\ \hline
$10^3$ & 11 & 43 & 13.629 & 7.537448 & 16 & 6.8\% & 148 & 999\\ \hline
$10^4$ & 12 & 163 & 58.5556 & 30.621093 & 33 & 2.73\% & 1526 & 9977\\ \hline
$10^5$ & 13 & 907 & 275.95915 & 177.355129 & 128 & 0.795\% & 12758 & 94680\\ \hline
$10^6$ & 16 & 4600 & 1357.37834 & 780.786047 & 4096 & 0.256\% & 235086 & 979501\\ \hline
\end{tabular}
$$
\caption{\label{table:misere}Statistics on the 
mis\`ere foreclosed \SP\ sequence}
\end{table}

Note that the growth of the mean of the foreclosed \SP\ numbers
is approximately logarithmic, which 
shows that even an {\em arithmetic period}~\cite[Chapter~4]{WW}
 cannot be expected on the considered interval.
Observe also the intriguing fact that the most 
frequently encountered foreclosed \SP\ number on the considered intervals
is always of the form $2^k$ or $2^k+1$ (which seems to be true for
every interval of type $[1,n]$).

\modif{
In fact, it appears that this foreclosed \SP\ sequence
is related to the \SP\ sequence of the octal game
{\bf 0.13337} under normal play by the relation
$F^-(P_n)=\rho_{0.13337}(H_{n-2})$, for every $n$, $n\ge 2$,
where $H_{n-2}$ denotes the heap of size $n-2$.
It is easy to check that this relation holds for paths $P_2$,
$P_3$ and $P_4$.
Now, let us write the options of $P_n$, $n\ge 5$, which corresponds
to $H_{n-2}$,
as follows:
$(i)$ $P_{n-2}$, which corresponds to $H_{n-4}$,
$(ii)$ $P_{n-3}$, which corresponds to $H_{n-5}$,
$(iii)$ $P_{n-4}\cup P_1 \simeq P_{n-4}$ (since $P_1$ is losing in one move),
which corresponds to $H_{n-6}$,
$(iv)$ $P_{n-5}\cup P_2 \simeq P_{n-5}$ (since $P_2$ is losing in one move),
which corresponds to $H_{n-7}$,
and $(v)$ $\{P_{n-5-j}\cup P_{2+j}$, $1\le j\le n-8\}$,
which corresponds to $\{H_{n-7-j}\cup H_j$, $1\le j\le n-8\}$.
Therefore, in terms of heaps, we get:
$(i)$ we can remove 2 elements in a heap, leaving 1 or 0 heaps,
$(ii)$ we can remove 3 elements in a heap, leaving 1 or 0 heaps,
$(iii)$ we can remove 4 elements in a heap, leaving 1 or 0 heaps,
$(iv)$ we can remove 5 elements in a heap, leaving 1 or 0 heaps,
and $(v)$ we can remove 5 elements in a heap, leaving 2 heaps.
Since we can remove 1 element only from a heap of size one, we
get exactly the rules of the octal game {\bf 0.13337}.
}

\modif{
Up to now, it is not known whether the \SP\ sequence
of this octal game is periodic or not~\cite{WEB-FLAMMENKAMP}.
}

\subsection{Conjunctive compound}

Recall that if $G=P_{i_1}\cup P_{i_2}\cup\dots\cup P_{i_k}$ is a graph made of $k$ disjoint
paths, we then have $O(G)=\{G_{i_1}, G_{i_2},\dots,G_{i_k}\}$ with
$G_{i_j}\in O(P_{i_j})$ for every $j$, $1\le j\le k$.

\modif{This version of our game is easy to solve. In both normal
and mis\`ere play, it can be checked that there are
only a finite number of (small) losing paths.
Therefore, 
we can easily determine the remoteness 
$R^+(P)$ (resp. $R^-(P)$) of any path $P$. 
}

\vskip 4mm

\centre{{\sc Normal Play}}

\noindent
Recall that if $O(G)=\{G_1,G_2,\dots,G_k\}$, the normal remoteness $R^+(G)$
of $G$ is given by:
$$\left\{\begin{array}{ll}
 R^+(G)=0 & \mbox{if\ } O(G)=\emptyset \\
 \\
 R^+(G)=1+min_{even}\{R^+(G_1),R^+(G_2),\dots,R^+(G_k)\} &
    \mbox{if $\exists\ j\in[1,k]$ s.t. $R^+(G_j)$ is even,}\\
    \\
 R^+(G)=1+max_{odd}\{R^+(G_1),R^+(G_2),\dots,R^+(G_k)\} &
   \mbox{otherwise.}
\end{array}
\right.$$

We prove the following:

\begin{theorem}
The normal remoteness $R^+$ of paths satisfies:\\
1.  $R^+(P_1)=R^+(P_2)=R^+(P_3)=1$,\\
2. $R^+(P_4)=R^+(P_5)=2$,\\
3. $R^+(P_6)=R^+(P_7)=R^+(P_8)=3$,\\
4. $R^+(P_9)=R^+(P_{10})=4$,\\
5. $R^+(P_n)=3$, for every $n\ge 11$.
\end{theorem}

\begin{proof}
The first four points can easily be checked.
Let now $n\geq 11$. Observe that $P_{n-7}\,\cup P_4\in O(P_n)$. 
By induction on $n$, and thanks to the remoteness of small paths, we have
$R^+(P_{n-7}\,\cup P_4)=min_{even}\{ R^+(P_{n-7}), R^+(P_4)\}=min_{even}\{ R^+(P_{n-7}), 2\}=2$
(since $n-7\ge 4$ we have $R^+(P_{n-7})\ge 2$).
Therefore, we get $R^+(P_n)=1+2=3$.
\end{proof}

We thus obtain:

\begin{corollary}
$\LL=\{0,4,5,9,10\}$.
\end{corollary}

Let now $G=P_{i_1}\cup P_{i_2}\cup\dots\cup P_{i_\ell}$ be any disjoint union
of paths and assume $i_1\le i_2\le\dots\le i_\ell$.
Clearly, the outcome of $G$ is $\PP$ if and only
if $i_1\in\{4,5,9,10\}$, which can be decided in linear time. 
Suppose now that $G$ is a $\NN$-position.
If $i_1\le 3$, one can win in one move.
If $6\le i_1\le 8$, one can play in such a way that $P_{i_1}$ gives a
path of length 4 or 5 and any other component gives a path of length
at least 4.
Finally, if $i_1\ge 11$, one can play in such a way that each component 
of order $p$ gives rise to $P_4\cup P_{p-7}$.
Finding such a winning move can thus be done in linear time.

\vskip 4mm

\centre{{\sc Mis\`ere play}}

\noindent
Similarly, if $O(G)=\{G_1,G_2,\dots,G_k\}$, the mis\`ere remoteness $R^-(G)$
of $G$ is given by:
$$\left\{\begin{array}{ll}
 R^-(G)=0 & \mbox{if\ } O(G)=\emptyset \\
 \\
 R^-(G)=1+min_{odd}\{R^-(G_1),R^-(G_2),\dots,R^-(G_k)\} &
    \mbox{if $\exists\ j\in[1,k]$ s.t. $R^-(G_j)$ is odd,}\\
    \\
 R^-(G)=1+max_{even}\{R^-(G_1),R^-(G_2),\dots,R^-(G_k)\} &
   \mbox{otherwise.}
\end{array}
\right.$$

We prove the following:

\begin{theorem}
The mis\`ere remoteness $R^-$ of paths satisfies:\\
1. $R^-(P_1)=R^-(P_2)=1$, \\
2. $R^-(P_n)=2$ for every $n\ge 2$.
\end{theorem}

\begin{proof}
The first point is obvious. Similarly, we can easily check that $R^-(P_3)=R^-(P_4)=2$.
Let now $n\geq 5$. 
Observe that $P_1\cup P_{n-4}\in O(P_n)$. 
By induction on $n$, and thanks to the remoteness of small paths, we have
$R^-(P_1\cup P_{n-4})=min_{odd}\{ R^-(P_1),R^-(P_{n-4})\}=
min_{odd}\{1,R^-(P_{n-4})\}=1$ (since $n-4>0$). 
Thus, we get $R^-(P_n)=1+1=2$.
\end{proof}

And therefore:

\begin{corollary}
$\LL=\{1,2\}$.
\end{corollary}

Hence, if $G$ is a disjoint union of paths, the outcome
of $G$ is $\PP$ if and only if the shortest component in $G$
has order 1 or 2, which can be decided in linear time. 
If $G$ is a $\NN$-position, a winning
move can be obtained, again in linear time, by playing for instance in
such a way that each component gives rise to a path of order 1.

\subsection{Continued conjunctive compound}

In this section, we will compute the suspense number 
$S^+(P_n)$ under normal play (resp. $S^-(P_n)$ under mis\`ere play)
for each path $P_n$.
Note that these two functions are {\em additive}~\cite[p.~177]{ONAG}
and we have $S^+(P_i\cup P_j)=max\{S^+(P_i),S^+(P_j)\}$
(resp. $S^-(P_i\cup P_j)=max\{S^-(P_i),S^-(P_j)\}$) for every
two paths $P_i$ and $P_j$.

\vskip 4mm

\centre{{\sc Normal play}}

\noindent
Recall that if $O(G)=\{G_1,G_2,\dots,G_k\}$, the normal suspense number $S^+(G)$
of $G$ is given by:
$$\left\{
\begin{array}{ll}
 S^+(G)=0 & \mbox{if\ } O(G)=\emptyset \\
 \\
 S^+(G)=1+max_{even}\{S^+(G_1),S^+(G_2),\dots,S^+(G_k)\} &
    \mbox{if $\exists\ j\in[1,k]$ s.t. $S^+(G_j)$ is even,}\\
    \\
 S^+(G)=1+min_{odd}\{S^+(G_1),S^+(G_2),\dots,S^+(G_k)\} &
   \mbox{otherwise.}
\end{array}
\right.$$

Then we prove the following:

\begin{theorem}
The normal suspense number $S^+$ of paths is an increasing
function and satisfies for every $n\ge 0$:\\
1.  $S^+(P_{5(2^n-1)})=2n$,\\
2.  $S^+(P_k)=2n+1$, for every $k\in [5(2^n-1)+1;5(2^{n+1}-1)-2]$,\\
3.  $S^+(P_{5(2^{n+1}-1)-1})=2n+2$.
\label{th:normal_suspense}
\end{theorem}

\begin{proof}
We proceed by induction on $n$.
For $n=0$, we can easily check that 
$S^+(P_0)=0$, $S^+(P_1)=S^+(P_2)=S^+(P_3)=1$ and that $S^+(P_4)=S^+(P_5)=2$.

Assume now that the result holds for every $p$, $0\le p<n$ and let
$k\in [5(2^n-1);5(2^{n+1}-1)-1]$. 
We consider three cases.

\begin{enumerate}
\item $k=5(2^n-1)$.\\
Since $\ceil{\frac{k-3}{2}}=5.2^{n-1}-4>5(2^{n-1}-1)$, 
using induction hypothesis, we get
$S^+(P_j)=2n-1$ for every $j$, $\ceil{\frac{k-3}{2}}\le j\le k-4$,
and thus $max(S^+(P_i),S^+(P_j))=2n-1$ for every $i,j$,
$j\ge i\ge 1$, $i+j=k-3$.
Therefore, since $S^+(P_{k-2})=S^+(P_{k-3})=2n-1$, $P_k$ has no option
with even suspense number and thus:
$$
\begin{array}{rcll}
S^+(P_k) & = & 1\ +\ min_{odd}\ ( & \{ S^+(P_{k-2}),S^+(P_{k-3})\} \\
  & & & \cup\ \{ max(S^+(P_i),S^+(P_j)),\ j\ge i\ge 1,\ i+j=k-3\}\ \ )\\
  & = & 1\ +\ min_{odd}\ ( & \{2n-1\}\
   \cup\ \{2n-1\}\ \ )\\
  & =& 2n
\end{array}
$$

\item $k\in [5(2^n-1)+1;5(2^{n+1}-1)-2]$.\\
Note first that for every such $k$,
$P_{5(2^n-1)}\cup P_{k-3-5(2^n-1)}$ is an option of $P_k$ with
even suspense number, since $k-3-5(2^n-1)\le 5(2^{n+1}-1)-2-3-5(2^n-1)
=5(2^n-1)-10<5(2^n-1)$ and, thus, 
$max(S^+(P_{5(2^n-1)}),S^+(P_{k-3-5(2^n-1)}))=S^+(P_{5(2^n-1)})=2n$
(thanks to the induction hypothesis and Case 1 above).
Therefore:
$$
\begin{array}{rcll}
S^+(P_k) & = & 1\ +\ max_{even}\ ( & \{ S^+(P_{k-2}),S^+(P_{k-3})\} \\
  & & & \cup\ \{ max(S^+(P_i),S^+(P_j)),\ j\ge i\ge 1,\ i+j=k-3\}\ \ ).
\end{array}
$$
We now proceed by induction on $k$.
We have
$$
\begin{array}{rcl}
S^+(P_{5(2^n-1)+1}) & = & 1\ +\ max_{even}\ (\ \ \{ S^+(P_{5(2^n-1)-1}),S^+(P_{5(2^n-1)-2})\} \\
  & &  \cup\ \{ max(S^+(P_i),S^+(P_j)),\ j\ge i\ge 1,\ i+j=5(2^n-1)-2\}\ \ )\\
  & = & 1\ +\ max_{even}\ (\ \{2n,2n-1\}\ \cup\ \{2n-1,2n\}\ \ )\\
  & = & 2n+1.
\end{array}
$$
and, similarly, $S^+(P_{5(2^n-1)+2})=S^+(P_{5(2^n-1)+3})=2n+1$.
Then, using induction hypothesis, we get
$$
\begin{array}{rcll}
S^+(P_{k}) & = & 1\ +\ max_{even}\ (& \{ S^+(P_{k-2}),S^+(P_{k-3})\} \\
  & & & \cup\ \{ max(S^+(P_i),S^+(P_j)),\ j\ge i\ge 1,\ i+j=k-3\}\ \ )\\
  & = & 1\ +\ max_{even}\ (& \{2n-1\}\ \cup\ \{2n-1,2n\}\ \ )\\
  & = & 2n+1.
\end{array}
$$

\item $k=5(2^{n+1}-1)-1$.\\
Thanks to Case 2 above, we have $S^+(P_{k-2})=S^+(P_{k-3})=2n+1$.
Moreover, 
since $\ceil{\frac{k-3}{2}}=5.2^{n}-3>5(2^{n}-1)$, 
using induction hypothesis and Case 2 above, we get
$S^+(P_j)=2n+1$ for every $j$, $\ceil{\frac{k-3}{2}}\le j\le k-4$,
and thus $max(S^+(P_i),S^+(P_j))=2n+1$ for every $i,j$,
$j\ge i\ge 1$, $i+j=k-3$.
Hence, $P_k$ has no option
with even suspense number and thus:
$$
\begin{array}{rcll}
S^+(P_k) & = & 1\ +\ min_{odd}\ ( & \{ S^+(P_{k-2}),S^+(P_{k-3})\} \\
  & & & \cup\ \{ max(S^+(P_i),S^+(P_j)),\ j\ge i\ge 1,\ i+j=k-3\}\ \ )\\
  & = & 1\ +\ min_{odd}\ ( & \{2n+1\}\
   \cup\ \{2n+1\}\ \ )\\
  & =& 2n+2
\end{array}
$$

\end{enumerate}
\end{proof}

And therefore:

\begin{corollary}
$\LL=\{5(2^n-1),\ n\ge 0\}\ \cup\ \{5(2^{n+1}-1)-1,\ n\ge 0\}$.
\end {corollary}

Note that Theorem~\ref{th:normal_suspense} shows that the normal
suspense sequence of paths has a {\em geometric period} with
geometric ratio 2.

Let $G=P_{i_1}\cup P_{i_2}\cup\dots\cup P_{i_\ell}$ be a disjoint union
of paths and assume $i_1\le i_2\le\dots\le i_\ell$.
The position $G$ has outcome $\PP$ if and only if $i_\ell\in\LL$, which
can be decided in linear time.
Now, if $G$ is a $\NN$-position,
let $r$ be the greatest integer such that $t=5(2^r-1)<i_\ell$.
A winning move can be obtained by playing in such a way that each component of
order $p>t$ gives rise to $P_{t-1}$ (if $p=t+1$), to $P_t$ (if $p=t+2$)
or to $P_t\cup P_{p-t-3}$ (otherwise). Such a move clearly leads to a
$\PP$-position and can be found in linear time.

\vskip 4mm

\centre{{\sc Mis\`ere play}}

\noindent
Recall that if $O(G)=\{G_1,G_2,\dots,G_k\}$, the mis\`ere suspense number $S^-(G)$
of $G$ is given by:
$$\left\{
\begin{array}{ll}
 S^-(G)=0 & \mbox{if\ } O(G)=\emptyset \\
 \\
 S^-(G)=1+max_{odd}\{S^-(G_1),S^-(G_2),\dots,S^-(G_k)\} &
    \mbox{if $\exists\ j\in[1,k]$ s.t. $S^-(G_j)$ is odd,}\\
    \\
 S^-(G)=1+min_{even}\{S^-(G_1),S^-(G_2),\dots,S^-(G_k)\} &
   \mbox{otherwise.}
\end{array}
\right.$$

Then we prove the following:

\begin{theorem}
The mis\`ere suspense number $S^-$ of paths is an increasing
function and satisfies for every $n\ge 0$:\\
1.  $\modif{S^-}(P_{7.2^n-6)})=2n+1$,\\
2.  $\modif{S^-}(P_{7.2^n-5)})=2n+1$,\\
3.  $\modif{S^-}(P_k)=2n+2$ for every $k$, $7.2^n-4\le k\le 7.2^{n+1}-7$.
\label{th:misere_suspense}
\end{theorem}

\begin{proof}
The proof is very similar to that of Theorem~\ref{th:normal_suspense} and we thus omit it.
\end{proof}

And therefore:

\begin{corollary}
$\LL=\{7.2^n-6,\ n\ge 0\}\ \cup\ \{7.2^n-5,\ n\ge 0\}$.
\end {corollary}

As in normal play, determining the outcome of a disjoint union
of paths or finding a winning move from a $\NN$-position can be done
in linear time.

\subsection{Selective compound}

With selective compound, each player may play on any number of components
(at least one). 
As seen in Section~\ref{sec:conway}, it is enough to know the outcome of each
component to decide the outcome of their (disjoint) union.
Therefore, we shall simply compute a boolean function $\sigma$, defined
by $\sigma(P)=1$ (resp. $\sigma(P)=0$) if and only if $o(P)=\NN$
(resp. $o(P)=\PP$) for every path $P$.

Then we have:

$$
\left\{\begin{array}{rll}
\sigma(G)= & 0\ \mbox{(normal) or\ } 
             1\ \mbox{(mis\`ere)} & \mbox{if\ } O(G) = \emptyset, \\
\sigma(G)=     & 1 - min\{\sigma(G'),\ G'\in O(G)\} & \mbox{otherwise}.
\end{array}
\right.
$$

The function $\sigma$ is additive, under
both normal and mis\`ere play, and we
have $\sigma(P_i\cup P_j)=\sigma(P_i)\vee\sigma(P_j)$ 
(boolean disjunction) for any
two non-empty paths $P_i$ and $P_j$.

We shall prove that the sequence 
$\sigma(P_0)\sigma(P_1)\sigma(P_2)\dots\sigma(P_{n-1})\sigma(P_n)\dots\ $
has period 5 under normal play and period 7 under mis\`ere play.

\vskip 4mm

\centre{{\sc Normal play}}

We prove the following:

\begin{theorem}
For every $n\geq 0$, we have:\\
1. $\sigma(P_{5n})=\sigma(P_{5n+4})=0$,\\
2. $\sigma(P_{5n+1})=\sigma(P_{5n+2})=\sigma(P_{5n+3})=1$.
\label{th:selective-normal}
\end{theorem}

\begin{proof}
We proceed by induction on $n$. 
For $n=0$, the result clearly holds.
Assume now that the result holds up to $n-1$. Then we have:

\begin{enumerate}

\item 
Recall that 
$O(P_{5n})=\{ P_{5n-2},P_{5n-3}\}\cup\{P_i\cup P_j,\
j\ge i\ge 1,\ i+j=5n-3\}$. Hence:
$$\begin{array}{rl}
\sigma(P_{5n}) & = 1 - min\{\sigma(P'),\ P'\in O(P_{5n})\} \\
 & = 1 - min\{1,\ 1,\ min_{j\ge i\ge 1,\ i+j=5n-3}\{\sigma(P_i)\vee\sigma(P_j)\}\ \} \\
 & = 1 - min\{1,\ 1,\ min_{j=5n-8,\dots,5n-4}\{\sigma(P_{5n-3-j})\vee\sigma(P_j)\}\ \} \\
 & = 1 - min\{1,\ 1,\ min\{0\vee 1,\ 0\vee 1,\ 1\vee 0,\ 1\vee 0,\
                1\vee 1,\ \}\ \} \\
 & = 1 - 1 \\
 & = 0
\end{array}$$
We can check in a similar way that $\sigma(P_{5n+4})=0$.

\item 
Since $\sigma(P_{5n})=\sigma(P_{5n-1})=0$,
$P_{5n-1}\in O(P_{5n+1})$, 
$P_{5n}\in O(P_{5n+2})$ and
$P_{5n}\in O(P_{5n+3})$, 
we have $\sigma(P_{5n+1})=\sigma(P_{5n+2})=\sigma(P_{5n+3})=1$.

\end{enumerate}
\end{proof}

And therefore:

\begin{corollary}
$\LL=\{5n,\ n\ge 0\}\cup\{5n+4,\ n\ge 0\}$.
\end {corollary}

Now, the outcome of a disjoint union of paths if $\PP$
if and only if each component $P$ is such that $\sigma(P)=0$,
which can be decided in linear time.
A winning move from a $\NN$-position can be obtained by playing
on each component $P$ with $\sigma(P)=1$ in such a way that
this component gives rise to a path $P'$ with $\sigma(P')=0$,
as explained in the proof of Theorem~\ref{th:selective-normal}.
Here again, such a move can be found in linear time.

\vskip 4mm

\centre{{\sc Mis\`ere play}}

We prove the following:

\begin{theorem} 
For every $n\geq 0$, we have:\\
1. $\sigma(P_{7n+1})=\sigma(P_{7n+2})=0$,\\
2. $\sigma(P_{7n+a})=1$, for every $a$, $3\le a\le 7$.
\end{theorem}

\begin{proof}
We proceed by induction on $n$.
For $n=0$, the result clearly holds.
Assume now that the result holds up to $n-1$. Then we have:

\begin{enumerate}

\item 
Recall that 
$O(P_{7n+1})=\{ P_{7n-1},P_{7n-2}\}\cup\{P_i\cup P_j,\
j\ge i\ge 1,\ i+j=7n-2\}$. Hence:
$$\begin{array}{rl}
\sigma(P_{7n+1}) & = 1 - min\{\sigma(P'),\ P'\in O(P_{7n+1})\} \\
 & = 1 - min\{1,\ 1,\ min_{j\ge i\ge 1,\ i+j=7n-2}\{\sigma(P_i)\vee\sigma(P_j)\}\ \} \\
 & = 1 - min\{1,\ 1,\ min_{j=7n-9,\dots,7n-3}\{\sigma(P_{7n-2-j})\vee\sigma(P_j)\}\ \} \\
 & = 1 - min\{1,\ 1,\ min\{1\vee 1,\ 1\vee 1,\ 1\vee 1,\ 0\vee 1,\
                0\vee 1,\ 1\vee 0,\ 1\vee 0\  \}\ \} \\
 & = 1 - 1 \\
 & = 0
\end{array}$$
We can check in a similar way that $\sigma(P_{7n+2})=0$.

\item 
Since $\sigma(P_{7n+1})=\sigma(P_{7n+2})=0$,
$P_{7n+1}\in O(P_{7n+3})$, 
$P_{7n+1}\in O(P_{7n+4})$ and
$P_{7n+2}\in O(P_{7n+5})$, 
we have $\sigma(P_{7n+3})=\sigma(P_{7n+4})=\sigma(P_{7n+5})=1$.

Now, observe that $P_{7n+2}\,\cup\, P_1\in O(P_{7n+6})$. Since 
$\sigma(P_{7n+2})=\sigma(P_1)=0$,
we have $\sigma(P_{7n+2}\cup P_1)=
\sigma(P_{7n+2})\vee\sigma(P_1)=0\vee 0=0$, which implies
$\sigma(P_{7n+6})=1$.

Similarly, since $P_{7n+2}\,\cup\, P_2\in O(P_{7n+7})$, we get
$\sigma(P_{7n+7})=1$.

\end{enumerate}
\end{proof}

And therefore:

\begin{corollary}
$\LL=\{7n+1,\ n\ge 0\}\cup\{7n+2,\ n\ge 0\}$.
\end {corollary}

As in normal play, determining the outcome of a disjoint union
of paths or finding a winning move from a $\NN$-position can be done
in linear time.

\subsection{Shortened selective compound}

We will use the same boolean function $\sigma$ as in
the previous subsection.
In both normal and mis\`ere play, we prove that
the corresponding sequence is periodic
with period 5.
\vskip 4mm

\centre{{\sc Normal play}}

As we have noted in Section~\ref{sec:conway} all positions have the same
outcome as in the selective compound. Therefore, we get from
the previous subsection:

\begin{corollary}
$\LL=\{5n,\ n\ge 0\}\cup\{5n+4,\ n\ge 0\}$.
\end {corollary}

The outcome of disjoint union of paths and winning moves are also similar.

\vskip 4mm

\centre{{\sc Mis\`ere play}}

On the other hand, selective compound and shortened selective
compound behave differently under mis\`ere play.
For instance, if $G$ is made of $k$ isolated vertices
($G=P_1\cup P_1\cup\dots\cup P_1$), with $k\ge 2$, then
$G$ is a $\PP$-position in selective compound and a
$\NN$-position in shortened selective compound.

As observed in~\cite[Chapter~14]{ONAG} the function $\sigma$
is not additive under mis\`ere play.
For instance, $\sigma(P_1)=0$ while  
$\sigma(P_1\cup\dots\cup P_1)=1$,
and $\sigma(P_4)=\sigma(P_5)=\sigma(P_8)=1$ while $\sigma(P_5\cup P_4)=0$
and $\sigma(P_8\cup P_4)=1$.

We first prove the following lemma which allows us to determine
$\sigma(G)$ for every position $G$ made of at least two
components (paths).

\begin{lemma}
Let $G=P_{i_1}\cup P_{i_2}\cup\dots\cup P_{i_\ell}$, with $\ell\ge 2$, and
let $\lambda_i(G)$, $0\le i\le 4$, be the number of paths in $G$ whose
order is congruent to $i$, modulo 5. Then,\\
$$\sigma(G)=0\ \ \mbox{if and only if} \ \ 
\lambda_1(G)+\lambda_2(G)+\lambda_3(G)=0.$$
\label{lemma:sigma}
\end{lemma}

\begin{proof}
We proceed by induction on the order $n$ of $G$.
The result clearly holds for $n=2$ (in that case, $G=P_1\cup P_1$
and $\sigma(G)=1$).
Suppose now that the result holds for every $p<n$.

Recall that
$O(P_k)=\{P_{k-2},P_{k-3}\}\cup\{P_i\cup P_j,\ j\ge i\ge 1,\ i+j=k-3\}$
for every path with $k$ vertices. 
Hence, if $k\equiv 0$ or $4$ \modu{5},
then every option of $P_k$
contains a path with order $m\equiv 1$, $2$ or $3$ \modu{5}.
Therefore, if $\lambda_1(G)+\lambda_2(G)+\lambda_3(G)=0$
then for every option $G'$ of $G$ we get
$\lambda_1(G')+\lambda_2(G')+\lambda_3(G')\neq 0$.
By induction hypothesis, that means $\sigma(G')=1$
for every option $G'$ of $G$, and thus $\sigma(G)=0$.

Suppose now that $\lambda_1(G)+\lambda_2(G)+\lambda_3(G)>0$.
Note that every path $P_k$ with $k\equiv 1$, $2$ or $3$ \modu{5},
has either an empty option (if $k\le 3$) or an option $P_{k'}$
with $k'\equiv 0$ or $4$ \modu{5} (by deleting 2 or 3 vertices
on one extremity of $P_k$). Therefore, by choosing such a move
for every path of $G$ of order $k\equiv 1$, $2$ or $3$ \modu{5},
we get an option $G'$ of $G$ with $\sigma(G')=0$ (by induction
hypothesis) and thus $\sigma(G)=1$.
\end{proof} 

We can now prove the following:

\begin{theorem}
The boolean function $\sigma$ satisfies:\\
1. $\sigma(P_1)=\sigma(P_2)=\sigma(P_8)=\sigma(P_9)=0$,\\
2. $\sigma(P_i)=1$ for every $i\in\{3,4,5,6,7,10,11,12,13,14\}$,\\
3. $\sigma(P_{5n})=\sigma(P_{5n+4})=0$ for every $n\ge 3$,\\
4. $\sigma(P_{5n+1})=\sigma(P_{5n+2})=\sigma(P_{5n+3})=1$ for every $n\ge 3$.
\label{th:shortened-selective-misere}
\end{theorem}

\begin{proof}
The first values can easily by checked.
For cases 3 and 4 we proceed by induction on $n$.

Since $P_{5n-1}\in O(P_{5n+1})$, $P_{5n}\in O(P_{5n+2})$,
$P_{5n}\in O(P_{5n+\modif{3}})$ and, by induction hypothesis,
$\sigma(P_{5n-1})=\sigma(P_{5n})=0$, we get
$\sigma(P_{5n+1})=\sigma(P_{5n+2})=\sigma(P_{5n+3})=1$.

Observe (as in the proof of Lemma~\ref{lemma:sigma}) that
 every option of $P_{5n}$ or $P_{5n+4}$
contains a path of order $m\equiv 1$, $2$ or $3$ \modu{5}.
Therefore, by Lemma~\ref{lemma:sigma}, every such option
is a winning position, and thus 
$\sigma(P_{5n})=\sigma(P_{5n+4})=0$.
\end{proof}

And therefore:

\begin{corollary}
$\LL=\{1,2,8,9\}\cup\{5n,\ n\ge 3\}\cup\{5n+4,\ n\ge 3\}$.
\end {corollary}

Now, the outcome of a disjoint union of paths has outcome $\PP$
if and only if the order of every component belongs to the set $\LL$,
which can be decided in linear time.
A winning move from a $\NN$-position can be obtained by playing
on every component of order $p\notin\LL$ as indicated in the
proof of Theorem~\ref{th:shortened-selective-misere}.
Such a winning move
can be found in linear time.

\modif{It is worth noting here that the set of losing
paths is the same as under normal play (and, thus, as in
the selective compound game under normal play), except for a few
small paths, namely $P_0, P_1, P_2, P_4, P_5, P_8, P_9, P_{10}$
and $P_{14}$. We do not have any explanation of this fact.}

\section{Discussion}
\label{sec:discussion}

\modif{
In this paper, we have solved ten versions of Conway's compound
\NK\ on paths by providing the set of losing positions of every
such game (see Table~\ref{table:summary} for a summary of these results). 
In each case, the outcome of any position can be
computed in linear time. The question of finding a
losing option from any winning position (which gives the winning
strategy) can as well be solved
in linear time. 
}

\begin{table}
$$
\begin{tabular}{|r|l|}
\hline
\bf Compound version & \bf Losing set $\LL$ \\
\hline
disj. comp., normal play & $\{0,4,8,14,19,24,28,34,38,42\}$\\ 
     & $\cup\ \{54+34i,58+34i,62+34i,72+34i,76+34i,\ i\ge 0\}$ \\ 
disj. comp., mis\`ere play & {\em unsolved}\\ 
dim. disj. comp., normal play & $\{0,4,5,9,10,14,28,50,54,98\}$ \\
dim. disj. comp., mis\`ere play & {\em unsolved}\\
conj. comp., normal play & $\{0,4,5,9,10\}$\\
conj. comp., mis\`ere play & $\{1,2\}$\\
cont. conj. comp., normal play & $\{5(2^n-1),\ n\ge 0\}\ \cup\ \{5(2^{n+1}-1)-1,\ n\ge 0\}$\\
cont. conj. comp., mis\`ere play & $\{7.2^n-6,\ n\ge 0\}\ \cup\ \{7.2^n-5,\ n\ge 0\}$\\
sel. comp., normal play & $\{5n,\ n\ge 0\}\cup\{5n+4,\ n\ge 0\}$\\
sel. comp., mis\`ere play & $\{7n+1,\ n\ge 0\}\cup\{7n+2,\ n\ge 0\}$\\
short. sel. comp., normal play &  $\{5n,\ n\ge 0\}\cup\{5n+4,\ n\ge 0\}$\\
short. sel. comp., mis\`ere play & $\{1,2,8,9\}\cup\{5n,\ n\ge 3\}\cup\{5n+4,\ n\ge 3\}$\\ \hline
\end{tabular}
$$
\caption{\label{table:summary}Losing positions for compound \NK\ on paths}
\end{table}

The first natural question is to complete \modif{our analysis}, 
by solving the diminished disjunctive
compound under mis\`ere play and, of course, the longstanding
open problem of disjunctive compound under mis\`ere play.

It would also be interesting to extend our results to other
graph families, such as stars, trees or outerplanar graphs
(we can solve for instance continued conjunctive compound
\NK\ on stars).
Note here that all our results trivially extend to cycles
since we have $O(C_n)=\{P_{n-3}\}$ for every cycle length
$n\ge 3$.

Stromquist and Ullman studied in~\cite{STROMQUIST-ULLMAN-1993}
the notion of {\em sequential compounds} of games. In such
a compound game $G\rightarrow H$, no player can play on $H$
while $G$ has not ended. They proposed as an open question to
consider the following compound game. Let $<$ be a partial order on games
and $G=G_1\cup G_2\cup\dots\cup G_k$ be a compound game. Then,
a player can play on component $G_i$ if and only if there is no
other component $G_j$ in $G$ with $G_j>G_i$. This idea can be
applied to \NK\ on paths by ordering the components according
to their length. (Note that this new rule makes sense only for
disjunctive and selective compounds).

Another variation could be to study \NK\ on {\em directed\,}
paths (paths with directed edges), where each player deletes
a vertex together to its out-neighbours.
Such a directed version of \NK\ on general graphs has been considered
in~\cite{DUCHENE-GRAVIER-MHALLA-08} (see also~\cite{PHD-DUCHENE}), 
under the name of
{\em universal domination game}.

Finally, inspired by the selective rule, we could also consider
a {\em selective} \NK\ game, where each player deletes a vertex
together with {\em some} of its neighbours.
Restricted to paths, this game corresponds to the
octal game {\bf 0.777}, still unsolved, and lies in some sense between
{\em Kayles} and {\em Dawson's chess}.

\bibliography{biblio_games}
\bibliographystyle{plain}

\end{document}